\documentclass[10pt,fleqn]{article}
\usepackage{amssymb}

\newcommand{\name}[1]{\begin{flushleft}
                       \LARGE \bf #1
                       \end{flushleft}\vspace{-3mm}}

\newcommand{\Author}[1]{\begin{flushleft}
                       \it #1 \end{flushleft}}

\newcommand{\Adress}[1]{\begin{flushleft}
                       \it #1 \end{flushleft}}

\newcommand{\be}{\begin{equation}}
\newcommand{\ee}{\end{equation}}
\newcommand{\ba}{\hspace*{-5pt}\begin{array}}
\newcommand{\ea}{\end{array}}
\newcommand{\p}{\partial}
\newcommand{\ds}{\displaystyle}

\newcommand{\pbf}[1]{\mbox{\mathversion{bold}$#1$}}

\begin{document}

\name{Poincar\'e-invariant equations\\ with
a rising mass spectrum}

\medskip

\noindent{published in {\it Lettere al Nuovo Cimento},  1975,  {\bf 14}, N~12, P. 435--438.}

\Author{Wilhelm I. FUSHCHYCH\par}

\Adress{Institute of Mathematics of the National Academy of
Sciences of Ukraine, \\ 3 Tereshchenkivska  Street, 01601 Kyiv-4,
UKRAINE}

\noindent {\tt URL:
http://www.imath.kiev.ua/\~{}appmath/wif.html\\ E-mail:
symmetry@imath.kiev.ua}

\bigskip

\noindent
In recent years many papers have been devoted to the construction of infinite-com\-po\-nent
wave equations to describe properly the spectrum of strongly interacting particles [1, 2].
As a rule, the derived equations have a number of pathological properties:
the unrealistic mass spectra, the appearence of spacelike solutions $(p_\mu^2<0)$,
the breakdown of causality etc.~[2].

In this note we shall construct, in the framework of relativistic quantum mechanics,
the Poinca\'re-invariant motion equations with realistic mass spectra.
These equations describe a system with mass spectra of the form
$m^2=a^2+b^2 s(s+1)$, where $a$ and $b$ are arbitrary parameters. Such
equations are obtained by a reduction of the motion equation for two particles to a
one-particle equation which describes the particle in various mass and spin states.
It we impose a certain condition on the wave function of the derived equation, such
an equation describes the free motion of a fixed-mass particle with arbitrary (but fixed) spin~$s$.

Let us consider the motion equation for two free particles with masses $m_1=m_2=m$
and spins $s_1$ and $s_2$ in the Thomas--Bakamjian--Foldy form~[3]
\be
i\frac{\p \Phi(t,\pbf{x},\pbf{\xi})}{\p t} =(P_a^2 +M^2)^{1/2} \Phi(t,\pbf{x}, \pbf{\xi}),
\ee
where
\[
P_a=p_a^{(1)} +p_a^{(2)}, \qquad M=2(m^2+\pbf{k}^2)^{1/2},
\]
$p_a^{(1)}$, $p_a^{(2)}$ are components of the momenta of the two particle,
$\pbf{k}$ the relative mo\-men\-tum,  $\pbf{x}$ the co-ordinate of the centre of mass,
$\pbf{\xi}$ is the relative co-ordinate.

On the manifold of solutions $\{\Phi\}$ of eq. (1) the generators of the Poincar\'e group
$P_{1,3}$ have the form
\be
\ba{l}
\ds P_0=(P_a^2+M^2)^{1/2}, \qquad P_a=p_a=-i\frac{\p}{\p x_a}, \quad a=1,2,3,
\vspace{2mm}\\
\ds J_{ab}=M_{ab}+L_{ab}, \qquad M_{ab}=x_ap_b-x_bp_a, \qquad
L_{ab}=m_{ab}+S_{ab},
\vspace{2mm}\\
\ds m_{ab} =\xi_a k_b -\xi_b k_a, \qquad S_{ab}=s_{ab}^{(1)}+s_{ab}^{(2)}, \qquad
[x_a,p_b]_-=i\delta_{ab},
\vspace{2mm}\\
\ds {}[\xi_a,k_b]_-=i\delta_{ab}, \qquad [\xi_a, p_b]_-=0,
\ea
\ee
where $s_{ab}^{(1)}$ and $s_{ab}^{(2)}$ are the spin matrices satisfying the Lie algebra of
the rotation group $O_3$.

Equations (1) is invariant with respect to algebra (2) since the condition
\be
\left[ i\frac{\p}{\p t} -(P_a^2+M^2)^{1/2} , J_{\mu\nu}\right]\Phi=0, \qquad
\mu=0,1,2,3,
\ee
is satisfied. In spherical co-ordinates the operator $\pbf{k}^2$ is
\be
\pbf{k}^2 =\frac{1}{\xi^2} \frac{\p}{\p \xi}\left(\xi^2 \frac{\p}{\p \xi}\right) +\frac{1}{\xi^2}
m_{ab}^2, \qquad \xi\equiv \pbf{\xi}^2=\xi_1^2 +\xi_2^2 +\xi_3^2,
\ee
where $m_{ab}$ is the square of the angular momentum with respect to the centre of mass.

Let us impose on the function $\Phi(t,\pbf{x},\xi,\theta,\varphi)$ the condition
\be
\frac{\p \Phi(t,\pbf{x},\theta,\varphi)}{\p \xi}=0.
\ee
This condition means that the wave function $\Phi$ constant on the sphere of radius
$r_0=\xi\equiv\sqrt{\xi^2}$ with respect to internal variables $\xi_1$, $\xi_2$, $\xi_3$.
If we take into account the condition (5), eq. (1) now becomes
\be
i\frac{\p \Phi(t,\pbf{x},\theta, \varphi)}{\p t} =\left(p_a^2 +4m^2 +\frac{4}{r_0^2} m_{ab}^2\right)^{1/2}
\Phi(t,\pbf{x},\theta,\varphi).
\ee

Equation (6) may yield the mass spectrum only for the bosons so that $m_{ab}$
should be replaced by $L_{ab}$. Having done this, we obtain the equation
\be
i\frac{\p \Phi(t,\pbf{x},\theta, \varphi)}{\p t} =\left(p_a^2 +4m^2 +\frac{4}{r_0^2} L_{ab}^2\right)^{1/2}
\Phi(t,\pbf{x},\theta,\varphi).
\ee
Equation (7) shows that the mass operator $M^2 =P_0^2-P_a^2$ has on the set
$\{\Phi(t,\pbf{x},\theta,\varphi)\}$ the discrete mass spectrum of the form
\be
M^2 \Phi=\left(4m^2+\frac{4}{r_0^2}L_{ab}^2\right) \Phi =\left\{ 4m^2 +\frac{4}{r_0^2}s(s+1)\right\}
\Phi,
\ee
where
\be
s=0,1,2,\ldots \qquad \mbox{if} \qquad L_{ab}=m_{ab}=\xi_a k_b -\xi_b k_a,
\ee
\be
s=\frac 12, \frac 32, \frac 52, \ldots \qquad \mbox{if} \qquad L_{ab}=\xi_a k_b -\xi_b k_a+S_{ab},
\ee
$S_{ab}=\sigma_c/2$, $\sigma_c$ are the $2\times 2$ Pauli matrices.

In the case (9) the operator $M^2$ has a simple spectrum. In the case (10) the spectrum of
$M^2$ is twofold degenerated. In the general case the measure of the degeneracy depends
on the dimension of the matrices $S_{ab}$ realizing representations of the group $O_3$.

If we suppose that the energy operator  $P_0$ can have both the positive and negative
spectrum, then for fermions (the spectum (10)) we find the equation
\be
\ba{l}
\ds p_0 \Phi(t,\pbf{x}, \theta, \varphi)=\gamma_0\left(p_a^2 +4m^2 +\frac{4}{r_0^2}
L_{ab}^2\right)^{1/2} \Phi(t,\pbf{x},\theta, \varphi),
\vspace{2mm}\\
\ds p_0=i\frac{\p}{\p t}, \qquad \gamma_0 =\left( \begin{array}{cc} 1 & 0\\ 0 & -1 \end{array}\right),
\ea
\ee
where $\Phi$ is the four-component wave function. The integro-differential equation (11) may
be written in the symmetrical form with respect to the operators $p_0$, $p_a$
if the transformation~[4] is carried out on it
\be
{\mathcal U}=\frac{1}{\sqrt{2}} \left(1+\frac{\gamma_0{\mathcal H}}{\sqrt{{\mathcal H}^2}}\right),
\quad {\mathcal H}=\gamma_0 \gamma_c p_c +\gamma_0 \gamma_4
\left(a^2 +b^2 L_{cd}^2\right)^{1/2},
\quad c,d=1,2,3,
\ee
where $\gamma_0$, $\gamma_c$, $\gamma_4$ are the $4\times 4$ Dirac matrices,
$a^2 =4m^2$, $b^2=4/r_0^2$. After the transformation (12), eq. (11) takes the form
\be
\ba{l}
\ds p_0\Psi(t,\pbf{x},\theta, \varphi)=\left\{\gamma_0 \gamma_c p_c +\gamma_0 \gamma_4
\left(a^2 +b^2L_{cd}^2\right)^{1/2}\right\} \Psi(t,\pbf{x}, \theta, \varphi),
\vspace{2mm}\\
\ds  \Psi={\mathcal U}\Phi.
\ea
\ee

We now summarize that eq. (7) describes a boson system with increasiftg mass spectrum
if the operator  $L_{ab}$  has the form (9). Equation (13) (or eq. (7)) describes a fermion
system with increasing mass spectrum if the operator  $L_{ab}$ has the form (10).

The four-component eq. (13) (or (7)) may be used for describing the free motion of
a particle of nonzero mass with arbitrary half-integer spin $s$. Indeed, to do this it is sufficient
to impose the Poincar\'e-invariant condition on the wave function $\Psi$,
picking up a fixed spin from the whole discrete spectrum (10).

This condition has the form
\be
\frac{1}{M^2} W_\mu W^\mu \Psi(t,\pbf{x}, \theta, \varphi)=L_{ab}^2\Psi(t,\pbf{x},\theta,\varphi)
=s(s+1)\Psi,
\ee
where
\be
W_\mu=\frac 12 \varepsilon_{\mu\nu\alpha\beta} P^\nu J^{\alpha\beta},
\ee
$s$ is an arbitrary but fixed number from the set (10).

Equations (7), (13) may be obtained in another way. Let us consider the equation
\be
i\frac{\p \Phi(t,x_1,x_2,\ldots, x_6)}{\p t} =\left(p_1^2 +p_2^2 +\cdots+p_6^2+
\varkappa^2\right)^{1/2}\Phi(t,x_1,x_2,\ldots, x_6),
\ee
where $p_k=-i(\p/\p x_k)$, $k=1,2,\ldots,6$, $\varkappa$ is a constant. The equation is
invariant under the generalized Poincar\'e group $P_{1,6}$~[5].

$P_{1,6}$ is the group of rotations and translations in $(1+6)$-dimensional Minkowski space.
Equation (16) is invariant with respect to the algebra~[5]
\be
\ba{l}
\ds P_0=p_0 =i\frac{\p}{\p t}, \qquad P_k =p_k =-i\frac{\p}{\p x_k}, \quad k=1,2,\ldots,6,
\vspace{2mm}\\
\ds J_{\mu\nu}=x_\mu p_\nu -x_\nu p_\mu+S_{\mu\nu}, \qquad \mu,\nu=0,1,2,\ldots,6.
\ea
\ee

Equation (16), together with the supplementary condition of the type (5), is equivalent to eq.~(7).
This may be shown by passing from the variables $x_4$, $x_5$, $x_6$
to the new variables $\xi$, $\theta$, $\varphi$. It is to be emphasized, however, that the
supplementary condition of the type~(5) breaks down the invariance with respect to the
whole group $P_{1,6}$ but conserves the invariance relative to its subgroup
$P_{1,3} \subset P_{1,6}$.

\smallskip

\noindent
{\bf Note 1.} On the set $\{\Phi\}$ besides the representations of the Poincar\'e algebra
$P_{1,3}$ (the external algebra), we may construct one more algebra of Poincar\'e $K_{1,3}$
(the internal algebra). The representation of the algebra $K_{1,3}$ has the following form:
\be
\ba{l}
\ds K_0=\frac 12 M, \qquad K_a=k_a =-i\frac{\p}{\p \xi_a}, \qquad L_{ab}=m_{ab}+S_{ab},
\vspace{2mm}\\
\ds m_{ab}=\xi_ak_b -\xi_b k_a, \qquad
L_{0a}=-\frac 12 (\xi_a K_0 +K_0\xi_a) -\frac{S_{ab} k_b}{K_0+m}.
\ea
\ee
This algebra describes an intrinsic relative motion of the two-particle system with respect to
the centre of mass. The algebra $P_{1,3}$ describes a motion of the centre of mass.
Equations~(7),~(13) are not invariant in respect to the whole algebra $K_{1,3}$.

\smallskip

\noindent
{\bf Note 2.} We note that the results obtained do not contradict the O'Raifeartaigh's theorem~[6]
since the operators (2) of the algebra $P_{1,3}$ together with the operators (18) of the algebra
$K_{1,3}$ form the infinite-dimensional Lie algebra.

\smallskip

\noindent
{\bf Note 3.} Equation (13) jointly with tin condition (14) for the case $s=\frac 12$
is equivalent to the ordinary four-component Dirac equation for the particle with the spin
$s=\frac 12$.

\medskip

\begin{enumerate}
\footnotesize

\item Majorana E., {\it Nuovo Cimento}, 1932, {\bf 9}, 355;\\
Nambu Y., {\it Prog. Theor. Phys. Suppl.}, 1966, {\bf 37-38}, 368;\\
Fronsdal C., {\it Phys. Rev.}, 1967, {\bf 156}, 1665; \\
Barut A.O., Corrigan D., Kleinert H., {\it Phys. Rev.}, 1968, {\bf 167}, 1527.

\item Chodos A., {\it Phys. Rev. D.}, 1970, {\bf 1}, 2973
(The reader will find an extensive list of further references in it).

\item Bakamjian B., Thomas L.H., {\it Phys. Rev.}, 1953, {\bf 92}, 1300;\\
Foldy L.L., {\it Phys. Rev.}, 1961, {\bf 122}, 289.

\item Fushchych W.I., {\it Lett. Nuovo Cimento}, 1974, {\bf 11}, 508, {\tt quant-ph/0206153}.

\item Fushchych W.I., Krivsky I.Yu., {\it Nucl. Phys. B}, 1968, {\bf 7}, 79, {\tt
quant-ph/0206057};\\ Fushchych W.I., Krivsky I.Yu., {\it Nucl.
Phys. B}, 1969, {\bf 14}, 573, {\tt quant-ph/0206047};\\ Fushchych
W.I., {\it Theor. Math. Phys.}, 1970, {\bf 4}, 360 (in Russian).

\item O'Raifeartaion L., {\it Phys. Rev.}, 1965, {\bf 14}, 575.
\end{enumerate}
\end{document}